%% file: 0PaperID3141.tex
\documentclass[sigconf]{acmart}

\usepackage{booktabs}
\usepackage{amsfonts}
\usepackage{mathrsfs}
\usepackage{graphicx}
\usepackage{multirow}
\usepackage{algorithm}
\usepackage{algpseudocode}
\usepackage{array}
\usepackage{float}
\usepackage{adjustbox}
\usepackage{comment}
\usepackage{enumitem}

\usepackage{color, xspace}

\AtBeginDocument{%
  }
    
\copyrightyear{2026}
\acmYear{2026}
\setcopyright{cc}
\setcctype{by}
\acmConference[WWW '26] {Proceedings of the ACM Web Conference 2026}{April 13--17, 2026}{Dubai, United Arab Emirates.}
\acmBooktitle{Proceedings of the ACM Web Conference 2026 (WWW '26), April 13--17, 2026, Dubai, United Arab Emirates}
\acmISBN{979-8-4007-2307-0/2026/04}
\acmDOI{10.1145/3774904.3792568}

\begin{document}

\title{Detecting Miscitation on the Scholarly Web through LLM‑Augmented Text-Rich Graph Learning}

\author{Huidong Wu}
\affiliation{
	\institution{Chinese Academy of Sciences \& \\  City University of Hong Kong}
	\country{Beijing, China}}
\email{wuhuidong21@mails.ucas.ac.cn}

\author{Haojia Xiang}
\affiliation{
	\institution{City University of Hong Kong}
	\country{Hong Kong, China}}
\email{haojxiang2-c@my.cityu.edu.hk}

\author{Jingtong Gao}
\affiliation{
	\institution{City University of Hong Kong}
	\country{Hong Kong, China}}
\email{jt.g@my.cityu.edu.hk}

\author{Xiangyu Zhao}
\authornote{Corresponding author}
\affiliation{
	\institution{City University of Hong Kong}
	\country{Hong Kong, China}}
\email{xianzhao@cityu.edu.hk}

\author{Dengsheng Wu}
\authornotemark[1]
\affiliation{
	\institution{Shenzhen University \& \\ Beijing Dongbi Data Technology}
	\country{Shenzhen, China}}
\email{wds@szu.edu.cn}

\author{Jianping Li}
\affiliation{
	\institution{University of Chinese Academy of Sciences}
	\country{Beijing, China}}
\email{ljp@ucas.ac.cn}

\begin{abstract}

Scholarly web is a vast network of knowledge connected by citations. However, this system is increasingly compromised by miscitation, where references do not support or even contradict the claims they are cited for. Current miscitation detection methods, which primarily rely on semantic similarity or network anomalies, struggle to capture the nuanced relationship between a citation's context and its place in the wider network. While large language models (LLMs) offer powerful capabilities in semantic reasoning for this task, their deployment is hindered by hallucination risks and high computational costs. In this work, we introduce LLM-Augmented Graph Learning-based Miscitation Detector (LAGMiD), a novel framework that leverages LLMs for deep semantic reasoning over citation graphs and distills this knowledge into graph neural networks (GNNs) for efficient and scalable miscitation detection. Specifically, LAGMiD introduces an evidence-chain reasoning mechanism, which uses chain-of-thought prompting, to perform multi-hop citation tracing and assess semantic fidelity. To reduce LLM inference costs, we design a knowledge distillation method aligning GNN embeddings with intermediate LLM reasoning states. A collaborative learning strategy further routes complex cases to the LLM while optimizing the GNN for structure-based generalization. Experiments on three real-world benchmarks show that LAGMiD achieves state-of-the-art miscitation detection with significantly reduced inference cost. The implementation code is available at: https://github.com/Applied-Machine-Learning-Lab/WWW2026\_LAGMiD. 


\end{abstract}




\maketitle
\input{1Introduction}

\input{3Preliminary}

\input{4Framework}

\input{5Experiments}
\input{2RelatedWork}

\input{6Conclusion}
\section*{Acknowledgements}
This research was supported by grants from the National Natural Science Foundation of China (Nos. T2293774, 72442024, 72022021, 62502404). This research was also partially supported by Hong Kong Research Grants Council (Research Impact Fund No.R1015-23, Collaborative Research Fund No.C1043-24GF, General Research Fund No.11218325), Institute of Digital Medicine of City University of Hong Kong (No.9229503), Huawei (Huawei Innovation Research Program), Tencent (Tencent Rhino-Bird Focused Research Program, Tencent University Cooperation Project), Alibaba (CCF-Alimama Tech Kangaroo Fund No. 2024002), Didi (CCF-Didi Gaia Scholars Research Fund), Kuaishou (CCF-Kuaishou Large Model Explorer Fund, Kuaishou University Cooperation Project), and Bytedance.


\bibliographystyle{ACM-Reference-Format}
\bibliography{9Reference}

\appendix
\input{7Appendix}

\end{document}

%% file: 1Introduction.tex

\section{Introduction}
\label{sec:introduction}

The scholarly web constitutes a vast, dynamic, and interconnected ecosystem of knowledge~\cite{vaughan2003scholarly, zhang2025chain, birkle2020web}, where citation functions as its foundational currency~\cite{petroni2023improving, chen2025ai4research}. Through citations, researchers contextualize their work, substantiate claims, and construct upon existing research~\cite{wu2024citation}. The integrity of citation network is, therefore, paramount to the advancement of science. However, this foundation is increasingly undermined by the pervasive issue of miscitation, a phenomenon wherein a referenced source fails to support, or even contradicts, the assertion it is cited to uphold~\cite{bornmann2025citation, cobb2024problem, peoples2024defensive, peoples2023burden}. Empirical estimates indicate that up to 25\% of citations in the general scientific literature contain inaccuracies and mislead academics~\cite{peoples2023burden}. Whether arising from unintentional oversight or deliberate rhetorical manipulation, miscitations propagate misinformation~\cite{kotiaho1999papers, cobb2024problem}, distort the outputs of academic search engines~\cite{wang2025unleashing, ziems2024measuring}, and ultimately erode the collective trust in the scientific record~\cite{bai2025revolutionizing}.

\begin{figure}[t]
	\centering
	\includegraphics[clip=true, trim= 175 170 165 130, width=0.9\linewidth]{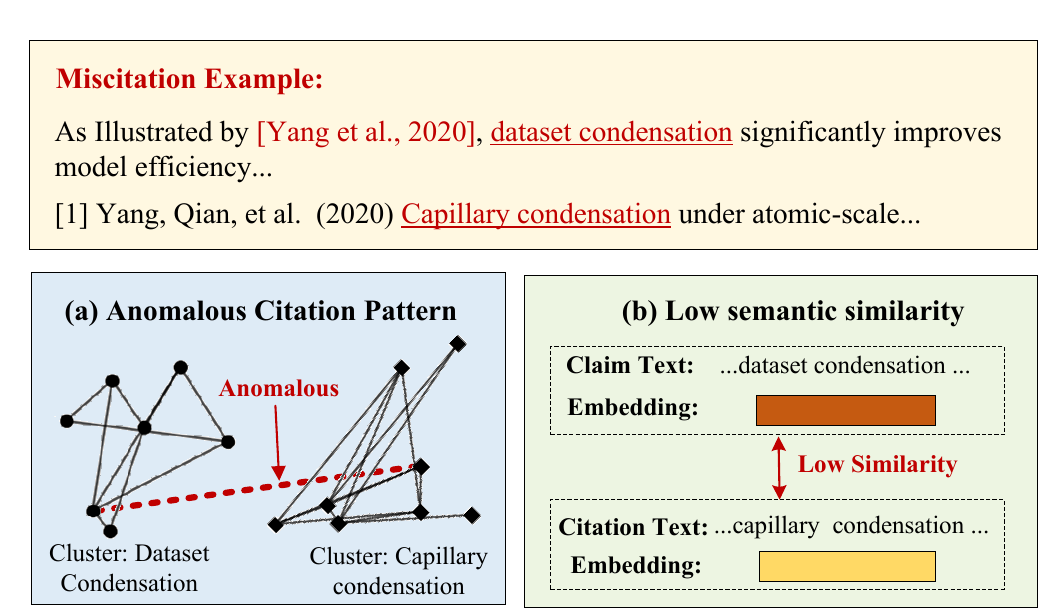}
	\caption{Illustrations of traditional miscitation detection methods. (a) Detection via anomalous citation patterns. (b) Detection via low semantic similarity.}
    \Description{Illustrations of traditional miscitation detection methods.}
	\label{Fig_tradiction}
	\vspace*{-5mm}
\end{figure}

The intrinsic graph nature of the scholarly web has motivated a substantial line of research to formulate miscitation detection as an edge classification task~\cite{liu2022deep, liu2024anomalous, kojaku2021detecting, wren2022detecting, pang2022editorial}. Early methods relied predominantly on network topology, identifying suspicious citations through structural anomalies~\cite{kojaku2021detecting,  wren2022detecting} such as atypical cross-disciplinary linkages (Figure~\ref{Fig_tradiction}(a)). While capable of revealing macro-level relational patterns, these approaches fundamentally overlook the semantic content of the citation context~\cite{liu2022deep}, which lies at the heart of citation validity. Subsequent efforts integrated local textual evidence by encoding features from citation sentences to augment classification~\cite{ren2023graph, liu2022deep, wadden2020fact}(Figure~\ref{Fig_tradiction}(b)). Despite these improvements, many models still operate on surface-level lexical similarities and lack the depth of semantic understanding required to distinguish strategically inserted or weakly grounded references.

The recent advent of large language models (LLMs) presents a transformative opportunity by offering profound semantic understanding and generative reasoning capabilities~\cite{boiko2023autonomous, zhang2025scientific}. In principle, LLMs can meticulously analyze the citing context against the cited paper's content, articulating rationales to assess relevance and accuracy~\cite{zhang2024detecting, ebadulla2025detecting, wang2025llm, shami2025fact, alvarez2024zero} . However, this powerful semantic analysis remains inherently local and computationally expensive. First, LLMs are susceptible to hallucinations, particularly when provided with incomplete or biased local context~\cite{ravichander2025halogen, lavrinovics2025knowledge}. Since they operate without awareness of the global citation network, they remain blind to systematic manipulation patterns such as when an author distorts the meaning of a reference or fabricates a supporting claim~\cite{deng2025next, press2024citeme}. Such structural anomalies can only be identified through a broader, network-level perspective~\cite{jin2024graph, lavrinovics2025knowledge}. These limitations highlight the need for a hybrid architecture that synergizes the deep semantic reasoning of LLMs with the global pattern recognition capabilities of graph models~\cite{xu2024multi}. Second, inference with large language models incurs substantial computational cost. As the volume of scientific publications grows exponentially, so does the number of references requiring analysis~\cite{wren2022detecting}. The sheer scale of the scholarly web, with its billions of citation edges, makes fine-grained, context-aware analysis using LLMs computationally intractable for the entire network~\cite{hu2025large, roy2025llm, pan2024distilling,li2024enhancing}. These challenges underscore the urgency of developing a robust and scalable method for miscitation detection.

In this study, we propose LAGMiD (\textbf{L}LM-\textbf{A}ugmented \textbf{G}raph Learning-based \textbf{Mi}scitation \textbf{D}etector), a novel framework that integrates the complementary strengths of LLM-based reasoning and graph-structured learning to identify miscitations across the scholarly web. To mitigate the risk of LLM hallucinations arising from limited local context, we introduce an evidence-chain reasoning mechanism powered by LLMs. This process leverages the inherent knowledge of the LLM to perform chain-of-thought reasoning~\cite{wei2022chain} over text-rich citation graphs by tracing reference sources. In other words, the judgment of a citation must be traced back to multi-hop citation sources. To overcome the substantial computational overhead of applying LLM reasoning at web scale, we propose distilling the LLM’s reasoning capability into a Graph Neural Network (GNN). The evidence-chain reasoning of the LLM aligns naturally with the message-passing mechanism of GNNs. Through a knowledge distillation strategy, we enable the GNN to internalize the LLM’s reasoning patterns, thereby achieving both faster inference and effective reasoning over graph-structured interaction data. Furthermore, given the diverse nature of miscitations where some manifest as anomalous structural patterns while others require deep semantic context analysis, we design a collaborative learning process that continuously aligns LLM knowledge with GNN structural representations. By iteratively refining GNN learning with LLM knowledge distillation across layers, our approach adaptively captures both structural and semantic anomalies, while simultaneously enhancing operational efficiency.

The main contributions of this paper are summarized as follows:

\begin{enumerate}[leftmargin=*]
\item  We identify key limitations in existing miscitation detection methods and propose LAGMiD. To the best of our knowledge, this is the first miscitation detection framework that integrates LLM reasoning with GNN structural modeling under a unified graph-learning paradigm.
\item We design an LLM-based evidence-chain mechanism using chain-of-thought prompting, and transfer this reasoning process to a student GNN via a novel knowledge distillation approach, enabling scalable and efficient inference. 
\item We introduce a collaborative learning strategy that iterat
ively aligns GNN representations with distilled LLM knowledge, allowing both models to fully exploit their respective strengths while enhancing overall inference efficiency.
\item We perform comprehensive experiments on three real-world datasets, demonstrating that LAGMiD consistently outperforms existing methods in miscitation detection.
\end{enumerate}

%% file: 3Preliminary.tex
\section{Preliminary}

\subsection{Text-Rich Citation Graph}

Scholarly webs naturally constitute a category of text-rich graphs (see Appendix~\ref{constructionTG} for construction details), where the node set comprises publication nodes and directed edges represent citations that denote referential relationships between nodes~\cite{chen2025ai4research, liu2020web}. Formally, we define a text-rich citation graph as 
\begin{equation}
\mathcal{G} = ( \mathcal{P}, \mathcal{E}, \mathcal{X}^{(P)}, \mathcal{X}^{(E)})
\end{equation}
where $\mathcal{P}$ denotes the set of publication nodes and $\mathcal{E} \subseteq \mathcal{P} \times \mathcal{P}$ represents the set of directed citation edges. Each node $p \in \mathcal{P}$ is associated with raw textual attributes $\mathcal{X}^{(P)}$, typically including the title, abstract, and/or full body of the publication. Similarly, each edge $e \in \mathcal{E}$ may be associated with textual attributes $\mathcal{X}^{(E)}$, such as citation contexts or explanatory claims extracted from the citing document. The raw texts in the citation graph can be encoded into dense vector representations via language model transformations:
\begin{equation}
X^{(P)} = f_{\text{LM}}(\mathcal{X}^{(P)}), \quad X^{(E)} = g_{\text{LM}}(\mathcal{X}^{(E)})
\end{equation}
where $f_{\text{LM}}$ and $g_{\text{LM}}$ denote suitable language model encoders for nodes and edges, respectively. 

\begin{figure*}[ht]
	\centering
	\includegraphics[clip=true, trim= 470 20 630 470, width=0.91\linewidth]{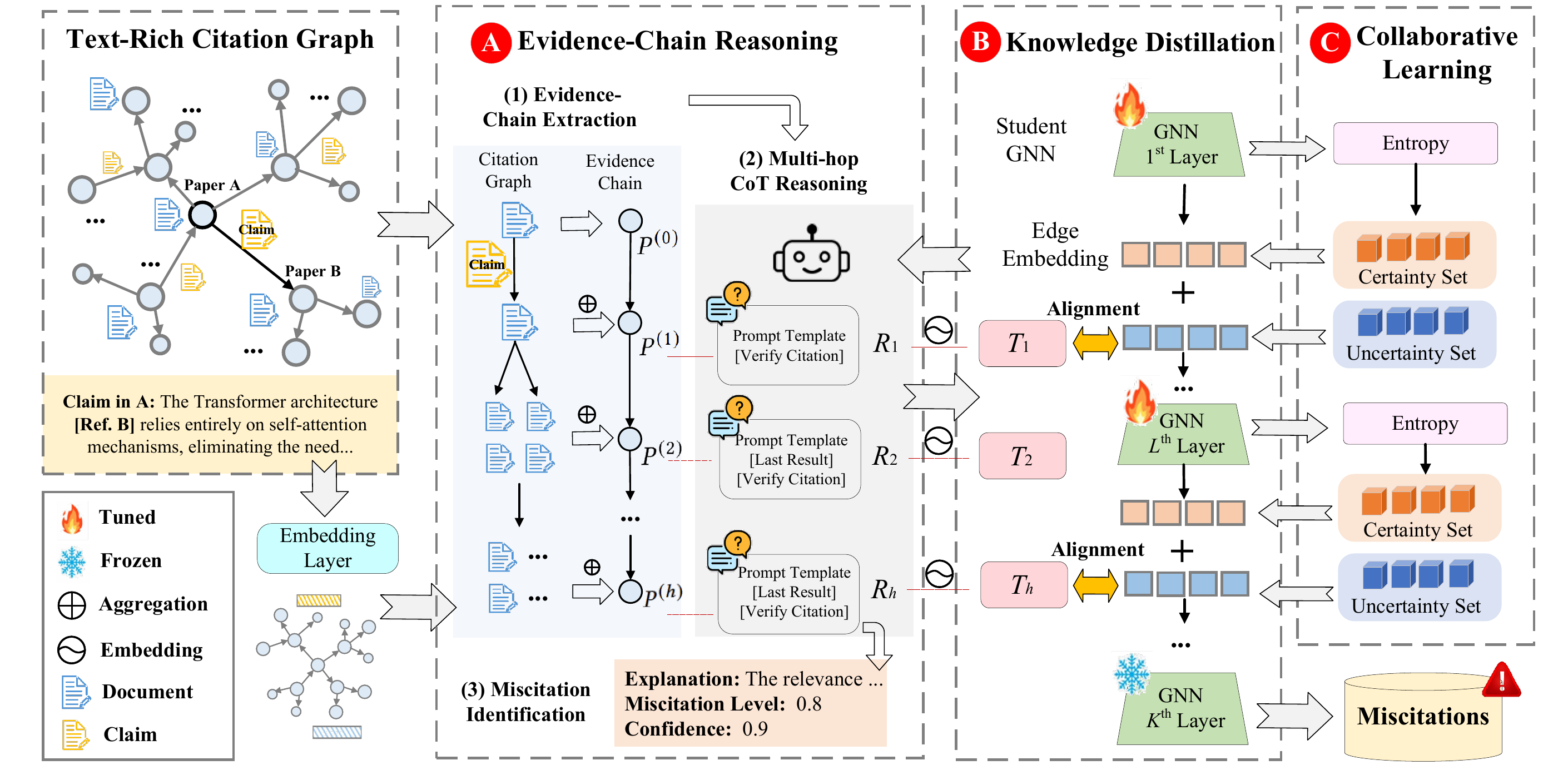}
	\caption{The overview of the proposed LAGMiD framework.}
    \Description{Overview Framework}
	\label{Fig_Overview}
	\vspace*{-3mm}
\end{figure*}

\subsection{Miscitation Detection Problem}

Building upon the formulation of a text-rich citation graph, we formalize the task of miscitation detection as a binary edge classification problem. Given a citation graph $\mathcal{G}$ and the annotated edge labels $\mathcal{Y} = \{0, 1\}$, where $y_{ij} = 1$ denotes a miscitation from paper $p_i$ to paper $p_j$ and $y_{ij} = 0$ a valid citation, the goal is to learn a binary classifier $f_\theta$ that assesses the semantic alignment between citing statements and cited documents. The model may incorporate multiple sources of evidence, including node features $\mathcal{X}^{(P)}$, edge features $\mathcal{X}^{(E)}$, and the structural properties of the global citation graph $\mathcal{G}$. The classification function is defined as:
\begin{equation}
f_\theta(e_{ij}) = \Psi\left( p_i, p_j, e_{ij}, \mathcal{G} \right) \rightarrow \{0, 1\}
\end{equation}
where $\Psi$ is a parameterized function that combines multiple evidence to output a binary prediction. Here, $f_\theta(e_{ij}) = 1$ indicates a miscitation, while $f_\theta(e_{ij}) = 0$ corresponds to a valid citation.

%% file: 4Framework.tex
\section{Methodology}

In this section, we introduce the proposed LLM-Augmented Graph Learning-based Miscitation Detector (LAGMiD). 
As illustrated in Figure~\ref{Fig_Overview}, LAGMiD is designed to leverage the complementary strengths of LLMs and GNNs to enhance the accuracy and robustness of miscitation detection. The architecture of LAGMiD is built around three tightly coupled components. First, it incorporates an LLM-driven evidence-chain reasoning mechanism that enables semantic analysis of citation validity by tracing reference sources through citation graphs (Figure~\ref{Fig_Overview}A). Second, it employs a knowledge distillation process that transfers the reasoning capabilities of LLMs into a more efficient GNN model, thereby facilitating scalable inference (Figure~\ref{Fig_Overview}B). Finally, LAGMiD adopts a collaborative learning strategy that jointly optimizes both the LLM and GNN components, enabling mutual enhancement through iterative interaction and shared objectives (Figure~\ref{Fig_Overview}C). The following subsections provide a detailed explanation of each component within the LAGMiD framework.

\subsection{LLM-based Evidence-Chain Reasoning}

Although LLMs exhibit a strong capacity for understanding semantic relationships between statements and their citation contexts, they are prone to hallucinations and misjudgments~\cite{ravichander2025halogen, lavrinovics2025knowledge}. These issues may manifest in the form of selective citation, misattribution, or distortion of the original conclusions. Passive reliance on direct statement-citation reasoning is inherently unstable. To address this, we introduce an active reasoning process termed Evidence-Chain Reasoning over the citation graph. The core mechanism of this approach requires the LLM to proactively traverse multi-hop citation chains, thereby validating the integrity of the evidence underlying a given claim. Leveraging the LLM’s chain-of-thought capabilities, our approach performs stepwise semantic reasoning across extracted citation chains. 

\vspace{0.5em}\noindent\textbf{Step 1: Evidence-Chain Extraction.}
We begin by extracting a multi-hop evidence chain from the text-rich citation graph $\mathcal{G} $, with the goal of tracing the source documents that support a given claim to enable systematic and verifiable reasoning. For a given citation edge $e_{ij} = (p_i, p_j) \in \mathcal{E}$, where paper $p_i$ cites paper $p_j$, we treat the statement text $\mathcal{X}^{(E)}_{ij}$ as the claim to be assessed. To identify supporting evidence, we construct a source-oriented subgraph rooted at $e_{ij}$, expanding along citation links for up to $K$ hops. Let $\mathcal{N}^+(p)$ denote the set of direct out-neighbors of node $p$, representing papers cited by $p$. The subgraph is generated iteratively as follows:
\begin{equation}
\mathcal{S}_1 = {p_j}, \quad
\mathcal{S}_{h} = \bigcup_{p \in \mathcal{S}_{h-1}} \mathcal{N}^+(p), \quad \text{for } h = 2, \ldots, K
\end{equation}
where $\mathcal{S}_h$ contains all documents reachable from $p_j$ within $h$ hops over the citation graph.

To ensure relevance and reduce redundancy, we apply a filtering strategy to $\mathcal{S}_h$ based on the vector representations $X^{(P)}$ and $X^{(E)}$ defined in the preliminary section. Let $X_{ij}^{(E)}$ represent the encoded form of the claim context, and $X_u^{(P)}$ the embedding of the textual content of a candidate paper $p_u$. Using a similarity function $\mathrm{sim}(X_{ij}^{(E)}, X_u^{(P)})$, such as cosine similarity, we select the top-$m$ most semantically relevant nodes $\mathcal{S}_h'$ from $\mathcal{S}_h$. Based on this filtered candidate set, we construct the evidence chain as a directed path:
\begin{equation}
\pi_{ij} = \left(P^{(0)}, P^{(1)},\ldots, P^{(h)}, \ldots,P^{(K)}\right)
\end{equation}
where $P^{(0)} = p_i$, $P^{(1)} = p_j$ and each $P^{(h)}$ is formed by aggregating the textual content $\mathcal{X}_u^{(P)}$ for all $p_u \in \mathcal{S}_h'$. This evidence chain serves as the basis for the subsequent reasoning procedure.

\vspace{0.5em}\noindent\textbf{Step 2: Multi-Hop Chain-of-Thought Reasoning.}
Given the extracted evidence chains $\pi_{ij}$, we employ a structured chain-of-thought reasoning process to perform stepwise verification across the multi-hop evidence trail. The goal of this stage is to determine whether each citation within the chain maintains semantic fidelity to the referenced content, thereby ensuring the consistent and accurate propagation of scholarly information.

At each hop $h$ ($1 \le h \le K$), the large language model is prompted to examine the relationship between two consecutive document sets, $P^{(h-1)}$ and $P^{(h)}$. The verification at hop $h$ evaluates whether the citation context $\mathcal{X}^{(E)}_{h-1,h}$ contained in $P^{(h-1)}$ faithfully represents the main claims or conclusions presented in the cited document $P^{(h)}$. Formally, the verification at hop $h$ is defined as:
\begin{equation}
V_h = \mathrm{LLM}\left(\phi_{\mathrm{verify}}\big(\mathcal{X}^{(P)}_{P^{(h)}}, \mathcal{X}^{(E)}_{h-1,h}, \mathcal{X}^{(P)}_{P^{(h-1)}}\big)\right)
\end{equation}
where $\phi_{\mathrm{verify}}$ denotes a task-specific prompting function that structures the verification input, and $V_h$ represents the verification outcome at hop $h$, indicating the semantic consistency.

The reasoning process for claim $e_{ij}$ evolves iteratively based on the verification results ${V_h}$. Each step integrates newly obtained evidence with the reasoning accumulated from previous hops. The reasoning state at hop $h$ is computed as:
\begin{equation}
R_h = \mathrm{LLM}\left(\phi_{\mathrm{cot}}\big(R_{h-1}, V_h, \mathcal{X}^{(P)}_{P^{(h)}}, \mathcal{X}^{(E)}_{h-1,h}\big)\right)
\end{equation}
where $R_h$ denotes the updated reasoning state after incorporating the result of hop $h$, and $\phi_{\mathrm{cot}}$ governs the evolution of the reasoning chain by combining the new verification with the accumulated contextual understanding.

The reasoning process continues sequentially until the terminal node $P^{(K-1)}$ is reached. The complete reasoning trajectory is expressed as  $\mathcal{R}_{ij} = [R_1, \ldots, R_{K}]$, which captures the progressive flow of evidence along the chain and documents the semantic integrity of each citation step. This cumulative reasoning trace provides an verifiable foundation for detecting potential miscitation or semantic distortion within the citation network. A detailed case study illustrating this reasoning process is provided in Appendix~\ref{case_study}.

\vspace{0.5em}\noindent\textbf{Step 3: Miscitation Identification.}
Building upon the reasoning trajectory $\mathcal{R}_{ij} = [R_1, \ldots, R_{K}]$ derived from the multi-hop verification process, the final step produces a structured assessment of the target citation edge $e_{ij}$. The LLM synthesizes the accumulated reasoning states and verification outcomes to generate a comprehensive miscitation judgment, formatted as a structured JSON object. This output comprises three core components: (1) a natural language explanation $U$ that summarizes the semantic (in)consistencies observed throughout the evidence chain; (2) a miscitation level $O$, represented as a normalized score in the range [0, 1], quantifying the severity of semantic misalignment, where higher values indicate more significant miscitations; and (3) a confidence score $C$, also in the range [0, 1], indicating the model's certainty in its assessment. Formally, the miscitation identification process is defined as:
\begin{equation}
\mathcal{M}_{ij}(U, O, C) = \mathrm{LLM}\left( \phi_{\mathrm{judge}}\left( \mathcal{R}_{ij}, \pi_{ij}, \mathcal{X}^{(E)}_{ij} \right) \right)
\end{equation}
where $\phi_{\mathrm{judge}}$ denotes a specialized prompt template that constrains the model to generate a JSON-structured output conforming to the specified schema. The resulting object $\mathcal{M}_{ij}$ provides both quantitative scoring and qualitative justification for citation quality, enabling transparent and interpretable miscitation detection that can be directly utilized in downstream analysis pipelines.

\subsection{LLM-to-GNN Knowledge Distillation}
While LLMs excel at evidence-chain reasoning over citation networks, their application to large-scale scholarly graphs remains constrained by high computational costs~\cite{hu2025large, roy2025llm, pan2024distilling,li2024enhancing}. To address this, we propose a knowledge distillation framework that transfers the reasoning capabilities of LLMs to a more efficient graph neural network (GNN). The inherent message-passing mechanism of GNNs offers a natural structural parallel to multi-hop evidence-chain reasoning structure of LLMs~\cite{yang2023empirical, yang2023mitigating}. Building on this, we introduce a knowledge distillation method that aligns LLM token embeddings with GNN node representations across hops. 

We begin by extracting structured knowledge representations from the evidence-chain reasoning process of the large language model (LLM). For each reasoning hop $h$ ($1 \leq h \leq K$), we obtain the LLM's hidden representations during the verification phase. Specifically, we extract token-level embeddings from the final transformer layer while the LLM processes the chain-of-thought prompt $\phi_{\mathrm{cot}}$ at hop $h$. Formally, for a reasoning state $\mathcal{R}_h$ at hop $h$, the LLM generates hidden representations as follows:
\begin{equation}
T_h = \mathrm{Transformer}(\mathcal{R}_h)
\end{equation}
where $T_h \in \mathbb{R}^{N_l}$, and $N_l$ denotes the sequence length at hop $h$. These representations capture the semantic understanding and intermediate reasoning state of the LLM at each step of the evidence-chain verification, providing crucial signals for miscitation detection.

Our student model is a graph neural network (GNN) that operates on the original citation graph $\mathcal{G}$, which includes node features $X^{(P)}$ and edge features $X^{(E)}$. We employ a $K$-layer GNN, where each layer updates node representations via neighborhood aggregation. At layer $k$, the node representation is updated according to:
\begin{equation}
z_u^{(k)} = \mathrm{UPDATE}\left(z_u^{(k-1)}, \mathrm{AGGREGATE}\left(\{z_v^{(k-1)} : v \in \mathcal{N}^+(u)\}\right)\right)
\end{equation}
where $z_u^{(0)} = X^{(P)}(p_u)$, and $\mathcal{N}^+(u)$ denotes the set of out-neighbors of node $u$. For edge-level representations between nodes $p_i$ and $p_j$, which are central to detecting miscitations, we compute:
\begin{equation}
e_{ij}^{(k)} = \mathrm{EDGE}\left(z_i^{(k)}, z_j^{(k)}, X^{(E)}_{ij}\right)
\end{equation}
Here, the $\mathrm{EDGE}$ function is instantiated as a neural network that combines node representations with corresponding edge features. Notably, the GNN’s $k$-th layer representations are aligned with the $k$-th hop in the LLM’s evidence chain, enabling a layer-wise correspondence between the two models.

To facilitate this alignment, we introduce a progressive distillation framework that encourages the GNN to mimic the LLM's hop-level reasoning representations. For each reasoning hop $h$, we minimize the distance between the LLM's representation $T_h$ and the GNN's corresponding edge representation $e_{h-1,h}^{(h)}$, which represents the citation edge between consecutive papers. We employ a learning objective based on the InfoNCE loss~\cite{wang2025infonce}:
\begin{equation}
\mathcal{L}_{\mathrm{distill}}^{(h)} = -\mathbb{E}\left[\log \frac{\exp(\mathrm{sim}(T_{h}, e_{h-1,h}^{(h)})/\delta)}{\sum_{j=1}^{N} \exp(\mathrm{sim}(T_{h}, e_{j}^{(h)})/\delta)}\right]
\label{distill}
\end{equation}
where $\mathrm{sim}(\cdot, \cdot)$ denotes cosine similarity, $\delta$ is a temperature hyperparameter, and negative samples are drawn from non-matching edge representations within the same training batch.
The overall distillation loss aggregates contributions across all reasoning hops:
\begin{equation}
\mathcal{L}_{\mathrm{distill}} = \sum_{h=1}^{K} \lambda_h \mathcal{L}_{\mathrm{distill}}^{(h)}
\end{equation}
Here, $\lambda_h$ denotes the layer-wise weighting coefficients, which are used to balance the contribution of each hop.

\subsection{Iterative Collaborative Learning Strategy}
While knowledge distillation from LLMs to GNNs provides a pathway for transferring semantic reasoning capabilities, applying this process uniformly across the entire training set is neither computationally efficient nor optimally effective. GNNs inherently excel at capturing structural dependencies within citation graphs and can resolve a substantial portion of miscitation cases through topological patterns alone~\cite{pang2025guard, liu2024anomalous, lin2024graph}. LLMs, conversely, offer superior performance in contextual interpretation and complex semantic reasoning~\cite{jin2024large, zhang2025notellm, wang2025rethinking, fu2023unified}. To leverage these complementary strengths, we design an iterative collaborative learning framework that facilitates selective, targeted distillation. 

The iterative collaboration begins with the GNN performing inference over the citation graph to generate node and edge representations. It then identifies regions of high predictive uncertainty, i.e., instances where the GNN itself exhibits low confidence and is likely to benefit from semantic verification of LLMs. For each citation edge $e_{ij}$, we quantify predictive uncertainty using the entropy of the GNN's output distribution:
\begin{equation}
U_{ij} = -\sum_{b=1}^{B} \hat{y}_{ij,b} \cdot \log(\hat{y}_{ij,b})
\end{equation}
where $\hat{y}_{ij,b}$ denotes the predicted probability for class $b$ in a binary classification setting ($B=2$). This entropy metric serves as a reliable indicator of the GNN's confidence and helps identify candidates for LLM-based refinement. To enable multi-hop alignment, uncertainty is evaluated at each GNN layer $k \in {1, 2, \dots, K}$. For each layer $k$, the top $w\%$ edges with the highest uncertainty values are selected to form a candidate set $\mathcal{E}_{\mathrm{uncertain}}^{(k)}$. 

For each uncertain edge $e_{ij} \in \mathcal{E}_{\mathrm{uncertain}}^{(k)}$, the LLM performs evidence-chain reasoning as described in Section 3.1. To ensure the quality of the knowledge being transferred, we apply a filtering mechanism: an edge is included in the distillation set $\mathcal{D}_{\mathrm{distill}}$ only if (1) the LLM's confidence score $C$ exceeds a predefined threshold $\tau_{\mathrm{conf}}$, and (2) the LLM's prediction aligns with the available ground-truth label. This selective filtering mitigates the risk of noise propagation and ensures that only high-confidence LLM reasoning guides the GNN's learning process. Targeted knowledge distillation is applied to the filtered set $\mathcal{D}_{\mathrm{distill}}$. The distillation loss is now specifically computed over this high-quality subset:
\begin{equation}
\mathcal{L}_{\mathrm{distill}}(\mathcal{D}_{\mathrm{distill}}) = \sum_{h=1}^{K} \lambda_h \mathcal{L}_{\mathrm{distill}}^{(h)}(\mathcal{D}_{\mathrm{distill}})
\end{equation}
where $\mathcal{L}_{\mathrm{distill}}^{(h)}(\mathcal{D}_{\mathrm{distill}})$ denotes the knowledge distillation loss at hop $h$ evaluated specifically over the edges in $\mathcal{D}_{\mathrm{distill}}$.

In parallel, the GNN is also optimized for the primary miscitation detection task via a supervised objective. The task-specific loss is defined as the binary cross-entropy over labeled edges:
\begin{equation}
\mathcal{L}_{\mathrm{task}} = -\frac{1}{|\mathcal{E}|} \sum_{e_{ij} \in \mathcal{E}} \left[ y_{ij} \log(\hat{y}_{ij}) + (1-y_{ij}) \log(1-\hat{y}_{ij}) \right]
\end{equation}
where $y_{ij}$ is the ground-truth label for edge $e_{ij}$, and $\hat{y}_{ij}$ is the GNN's predicted probability of miscitation. The total training objective combines both losses:
\begin{equation}
\mathcal{L}_{\mathrm{total}} = \beta_1 \mathcal{L}_{\mathrm{distill}}(\mathcal{D}_{\mathrm{distill}}) + \beta_2 \mathcal{L}_{\mathrm{task}}
\end{equation}
where $\beta_1$ and $\beta_2$ balance the contributions of targeted knowledge distillation and task-specific supervision.

%% file: 5Experiments.tex
\section{Experiments}
\label{sec:experiment}

\subsection{Experimental Setup}

\subsubsection{\textbf{Datasets}}
We evaluate the proposed LAGMiD framework on three real-world scholarly benchmarks: RED (Reference Error Detection)~\cite{zhang2024detecting}, SciFact~\cite{wadden2020fact, wadden2022scifact}, and S2ORC~\cite{lo2020s2orc}. The RED dataset is a human-annotated corpus designed for reference error detection, comprising statement–citation pairs extracted from journal articles across diverse domains including biomedical science, chemistry/materials, physics, and social sciences. SciFact serves as a benchmark for scientific claim verification, featuring expert-authored claims paired with relevant abstract sentences, along with corresponding labels and rationale spans. S2ORC constitutes a large-scale English-language academic paper corpus that furnishes comprehensive metadata and citation contexts, making it particularly suitable for miscitation analysis. Following Liu et al.~\cite{liu2024anomalous}, we construct a computer science subset of S2ORC for miscitation detection. The detailed preprocessing of all three datasets is provided in Appendix~\ref{dataset}, and Table~\ref{tab:dataset_overview} reports their statistics. 

\begin{table}[htbp]
\caption{Dataset statistics.}
\vspace*{-3mm}
\begin{adjustbox}{max width=\linewidth}
\begin{tabular}{lccc}
\hline
\textbf{Statistics} & \textbf{RED} & \textbf{SciFact} & \textbf{S2ORC} \\
\hline
\# Statement-Citation Pairs & 250 & 1,109 & 3,561 \\
\quad \# Miscitations & 126 & 463 & 936 \\
\quad \# Valid Citations & 124 & 646 & 2,625 \\
\# AvgLen    &  202   & 244 & 341 \\
\# MaxLen    &  506   & 1,085 & 1,714 \\
\# Train  & 175    &  809 & 2,493 \\
\# Valid & 25   & 150 & 356 \\
\# Test  & 50     & 150 & 712 \\
\hline
\# Nodes of Citation Graph   &  5,844   & 667 & 413 \\
\# Edges of Citation Graph    &  5,781   & 1,559 & 7,122 \\
\hline
\end{tabular}
\label{tab:dataset_overview}
\end{adjustbox}
\end{table}

\begin{table*}[t]
\caption{Overall performance comparison for miscitation detection on the three benchmark datasets. The superior result for each metric is highlighted in \textbf{bold}, and the second-best is \underline{underlined}. A * denotes cases where the result is statistically significantly better (\textit{p} < 0.05) than the second-best.}
\vspace*{-3mm}
\begin{adjustbox}{width=\textwidth}
\begin{tabular}{llccccccccc}
\toprule
\textbf{Dataset} & \textbf{Metric} & \textbf{GCN} & \textbf{GLAD} & \textbf{RoBERTa} & \textbf{SciBERT} & \textbf{GLM} & \textbf{Qwen} & \textbf{AnomalyLLM} & \textbf{GuARD} & \textbf{LAGMiD (Ours)} \\
\midrule
\multirow{3}{*}{RED}
& AUC        & 0.7559 & 0.7774 & 0.7068 & 0.7651 & 0.8220 & 0.8362 & 0.8982 & \underline{0.9100} & \textbf{0.9615}* \\
& F1         & 0.6857 & 0.6936 & 0.6082 & 0.7471 & 0.7202 & 0.8056 & 0.8354 & \underline{0.8571} & \textbf{0.9167}* \\
& Precision  & 0.7557 & 0.7774 & 0.5982 & 0.7182 & 0.8051 & 0.7682 & 0.8327 & \underline{0.8753} & \textbf{0.9167}* \\
\midrule
\multirow{3}{*}{SciFact}
& AUC        & 0.6045 & 0.6244 & 0.5677 & 0.6211 & 0.7909 & 0.7832 & \underline{0.8024} & 0.7906 & \textbf{0.8208}* \\
& F1         & 0.5138 & 0.6006 & 0.5252 & 0.6622 & 0.7242 & 0.7213 & \underline{0.7958} & 0.7714 & \textbf{0.8205}* \\
& Precision  & 0.6673 & 0.6677 & 0.6360 & 0.7102 & 0.7250 & \underline{0.7536} & 0.7485 & 0.7453 & \textbf{0.7724}* \\
\midrule
\multirow{3}{*}{S2ORC}
& AUC        & 0.6971 & \underline{0.7883} & 0.7459 & 0.7024 & 0.7129 & 0.7506 & 0.7547 & 0.7528 & \textbf{0.8100}* \\
& F1         & 0.6871 & \underline{0.7932} & 0.6330 & 0.7094 & 0.7621 & 0.7678 & 0.7381 & 0.7687 & \textbf{0.8256}* \\
& Precision  & 0.6606 & 0.7733 & 0.7081 & 0.7653 & 0.7681 & 0.7324 & 0.7420 & \underline{0.7740} & \textbf{0.8419}* \\
\bottomrule
\end{tabular}
\label{tab:overall_performance}
\end{adjustbox}
\end{table*}

\subsubsection{\textbf{Evaluation Metrics}}
For the binary classification task of miscitation detection, we employ three standard evaluation metrics: AUC, F1 score, and Precision. AUC (Area Under the ROC Curve) measures the model's overall ranking capability across all classification thresholds. Precision specifically quantifies the model's accuracy in identifying true miscitations among all predicted positives, which is particularly important for minimizing false alarms in practical applications. F1 score, as the harmonic mean of precision and recall, provides a balanced assessment of model performance. 

\subsubsection{\textbf{Implementation Details}}
Our implementation of LAGMiD employs Qwen3-8B as the backbone LLM for evidence-chain reasoning. For the GNN component, we implement a 2-layer GCN architecture with both hidden dimensions of 1024. The edge representation function $\mathrm{EDGE}(\cdot)$ is implemented as a 2-layer MLP that concatenates node representations with edge features. Node and edge textual features are encoded using SciBERT. 

We set the evidence chain length $K = 2$ with top-$m = 10$ nodes selected at each hop using cosine similarity filtering. In the collaborative learning framework, we set the confidence threshold $\tau_{\mathrm{conf}} = 0.7$ and select the top $w = 20\%$ most uncertain edges for LLM refinement at each iteration. Ohter hyperparameters including the distillation temperature $\delta$,  layer weights $\lambda_h$, and loss weighting coefficients $\beta_1$, $\beta_2$ are determined through grid search. The model is trained using the AdamW optimizer with an initial learning rate of 0.001 and weight decay of 0.01. We employ a batch size of 32 and train for 200 epochs with early stopping based on validation performance. All experiments are conducted on NVIDIA A100 GPU.

\subsubsection{\textbf{Baselines}}
We evaluate LAGMiD against eight baseline models spanning four categories:
\textbf{(1) GNN-based Methods.} This category includes two models: a standard Graph Convolutional Network (GCN)~\cite{kipf2016semi}, which serves as a fundamental graph-based baseline; and GLAD~\cite{liu2022deep}, a GNN that incorporates edge features specifically designed for anomalous citation detection.
\textbf{(2) Pretrained Language Model (PLM)-based Methods.} We consider two widely-used PLMs: RoBERTa~\cite{liu2019roberta}, an optimized Transformer encoder commonly employed as a strong text-based baseline; and SciBERT~\cite{beltagy2019scibert}, a domain-specific language model pretrained on scientific corpora.
\textbf{(3) Large Language Model (LLM)-based Methods.} This group comprises GLM4-9B~\cite{glm2024chatglm} and Qwen3-8B~\cite{yang2025qwen3}, two large-scale generative language models capable of capturing complex semantic patterns, making them well-suited for identifying anomalous citations.
\textbf{(4) Text-Rich Graph Learning Methods.} We also include advanced frameworks designed for text-rich graph learning, which are naturally transferable to miscitation detection. AnomalyLLM~\cite{liu2024anomalyllm} leverages LLMs to classify anomalous graph edges by integrating textual and structural information. GuARD~\cite{pang2025guard} performs anomaly detection by aligning rich textual content with graph structure within a unified framework, utilizing a large model as its backbone. 
Implementation details for all baseline models are provided in Appendix~\ref{app:baseline_details}.

\subsection{Overall Performance}

Table~\ref{tab:overall_performance} presents the overall performance of our proposed LAGMiD method compared to baseline approaches. Across all evaluated datasets and performance metrics, LAGMiD consistently achieves superior performance. Traditional GNN-based and PLM-based methods, which rely primarily on either structural features or semantic similarity, exhibit the weakest performance. This underscores their limited capability to effectively integrate structural perception with semantic understanding, both of which are critical for accurate miscitation detection. LLM-based methods, which leverage extensive pre-trained knowledge and reasoning capabilities, achieve improved results over traditional models. Nonetheless, they still fall short compared to text-rich graph learning approaches that more effectively integrate graph structural information with semantic reasoning. Among these, our LAGMiD significantly outperforms baselines including AnomalyLLM and GuARD. This performance gap suggests that general-purpose graph anomaly detection techniques may face limitations when adapted to the specific challenges of miscitation detection. Overall, the strong performance of LAGMiD highlights its effectiveness and robustness in addressing the nuanced requirements of miscitation tasks.

\subsection{Ablation Study}

To rigorously evaluate the contribution of each individual component within the proposed LAGMiD framework, we conduct a comprehensive ablation study. Specifically, we design four model variants, each constructed by removing a particular module:
\begin{itemize}[leftmargin=*]
    \item \textbf{w/o EC}: Removes evidence chain reasoning, relying solely on direct statement-citation reasoning performed by the LLM .
    \item \textbf{w/o KD}: Replaces knowledge distillation module by a naive feature concatenation strategy.
    \item \textbf{w/o LD}: Removes layer-wise distillation component and performs distillation only at the final representation layer.
    \item \textbf{w/o TD}: Removes targeted distillation strategy based on uncertainty sets, instead applying knowledge distillation uniformly across the entire training dataset.
\end{itemize}
The results of the ablation study, presented in Table~\ref{tab:ablation}, demonstrate that each component of our model contributes significantly to overall performance. Notably, the removal of the evidence chain reasoning module results in a substantial performance drop, underscoring the critical role of multi-hop reasoning in capturing complex citation dependencies. The knowledge distillation mechanism also offers marked improvements over the feature concatenation baseline, emphasizing the effectiveness of aligning the semantic representations between LLMs and GNNs. In addition, the layer-wise distillation strategy consistently enhances performance, suggesting that leveraging reasoning patterns across all GNN layers improves the overall representational quality. Lastly, the targeted distillation approach contributes further performance gains by directing the model’s focus toward more uncertain and informative samples, thereby optimizing the learning process.

\begin{table}[t]
\centering
\caption{Ablation study on different components of LAGMiD.}
\vspace*{-3mm}
\begin{adjustbox}{max width=\linewidth}
\begin{tabular}{lcccccc}
\toprule
\multirow{2}{*}{Model} &
\multicolumn{3}{c}{RED} &
\multicolumn{3}{c}{S2ORC} \\
\cmidrule(lr){2-4} \cmidrule(lr){5-7}
& AUC & F1 & Precision & AUC & F1 & Precision \\
\midrule
w/o EC & 0.9295 & 0.8412 & 0.8571 & 0.7480 & 0.7855 & 0.7728 \\
w/o KD & 0.9487 & 0.8511 & 0.8696 & 0.7586 & 0.7935 & 0.7885 \\
w/o LD & 0.9503 & 0.8696 & 0.8571 & 0.7700 & 0.8207 & 0.8035 \\
w/o TD & 0.9487 & 0.8511 & 0.8696 & 0.7625 & 0.7968 & 0.8277 \\
\hline
LAGMiD & \textbf{0.9615} & \textbf{0.9167} & \textbf{0.9167} & \textbf{0.8100} & \textbf{0.8256} & \textbf{0.8419} \\
\bottomrule
\end{tabular}
\end{adjustbox}
\label{tab:ablation}
\end{table}

\subsection{Effectiveness of Evidence-Chain Reasoning}
\label{subsec:evidence_chain_analysis}

To validate the effectiveness of our evidence-chain reasoning mechanism, we conduct ablation studies on three configurations: (1) \textbf{Directed}: without evidence-chain reasoning, using only direct statement-citation analysis; (2) \textbf{Unfiltered}: with evidence chains but without semantic filtering; and  (3) \textbf{Full}: the complete approach with semantic filtering. Experiments are conducted with two LLM backbones (Qwen3-8B and GLM4-9B).

\begin{table}[t]
\centering
\caption{Performance comparison of evidence-chain reasoning variants on different foundation models.}
\vspace*{-3mm}
\label{tab:evidence_chain_ablation}
\begin{adjustbox}{max width=\linewidth}
\begin{tabular}{llcccccc}
\toprule
\multirow{2}{*}{LLM} & \multirow{2}{*}{Model} &
\multicolumn{3}{c}{RED} & \multicolumn{3}{c}{S2ORC} \\
\cmidrule(lr){3-5} \cmidrule(lr){6-8}
 & & AUC & F1  & Precision & AUC & F1  & Precision \\
\midrule
\multirow{3}{*}{Qwen}
 & Directed         & 0.8362 & 0.8056 & 0.7682      & 0.7506 & 0.7678    & 0.7324 \\
 & Unfiltered  & 0.8486 & 0.8195 & 0.7725     & 0.7429 & 0.7738     & 0.7527 \\
 & Full        & \textbf{0.8543} & \textbf{0.8333} & \textbf{0.7857}      & \textbf{0.7680} & \textbf{0.7855}      & \textbf{0.7728} \\
\midrule
\multirow{3}{*}{GLM}
 & Directed         & 0.8220 & 0.7202 & 0.8051    & 0.7129 & 0.7621      & 0.7681 \\
 & Unfiltered  & 0.8267 & 0.7329 & 0.8100     & 0.7109 & 0.6835       & 0.6654 \\
 & Full        & \textbf{0.8385} & \textbf{0.7487} & \textbf{0.8182}      & \textbf{0.7504} & \textbf{0.7821}      & \textbf{0.7981} \\
\bottomrule
\end{tabular}
\end{adjustbox}
\end{table}

As shown in Table~\ref{tab:evidence_chain_ablation}, the Full configuration consistently outperforms all other variants. The performance gap between Unfiltered and Directed indicates that simply aggregating all multi-hop evidence does not reliably improve detection. In contrast, the superior performance of Full over Unfiltered underscores the importance of semantic filtering in suppressing noise. These findings collectively demonstrate that multi-hop reasoning, when coupled with semantic filtering, significantly enhances miscitation detection by enabling more accurate and contextually-grounded inference.

\subsection{Effectiveness of Knowledge Distillation}

\begin{figure}[t]
	\centering
	\includegraphics[clip=true, width=0.9\linewidth]{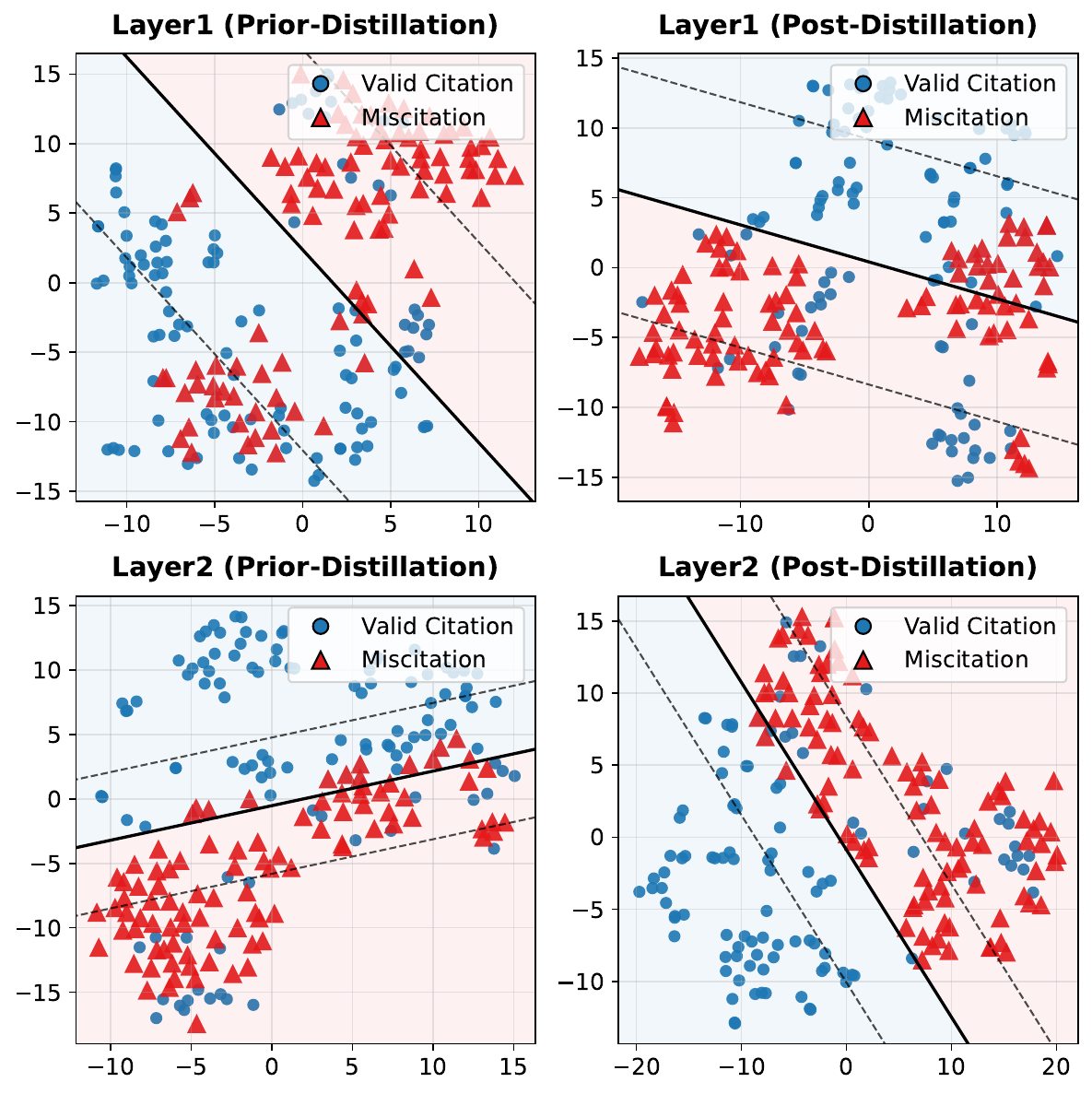}
	\caption{The t-SNE visualizations of citation embeddings from the RED, comparing representations before and after knowledge distillation across different GNN layers. Blue circles denote valid citations, while red triangles represent miscitations. The solid line indicates the decision boundary.}
    \Description{TSNE}
	\label{Fig_tsne}
	\vspace*{-5mm}
\end{figure}

We evaluate the effectiveness of our knowledge distillation framework by visualizing embeddings with t-SNE. As illustrated in Figure~\ref{Fig_tsne}, embeddings from two GNN layers are projected into a two-dimensional space, both before and after distillation. The visualizations reveal that each distillation step enhances the discriminability of GNN embeddings for miscitation detection. Prior to distillation, although some miscitations are identifiable, substantial overlap remains between miscitations and legitimate citations. After distillation, ambiguous instances shift closer to the decision boundary, resulting in improved class separation. The second-layer embeddings also exhibit a noticeable improvement, with a more distinct boundary emerging between the two classes.

These findings further validate the effectiveness of our layer-wise distillation strategy. Specifically, after the first-layer distillation and prior to the second-layer distillation, the GNN leveraging semantic information distilled from the LLM demonstrated enhanced classification performance by transferring structural cues. This indicates that the synergistic integration of the LLM and GNN components within the LAGMiD framework effectively facilitates more discriminative representation learning.

\subsection{Model Efficiency}

\begin{figure}[t]
	\centering
	\includegraphics[clip=true, width=0.9\linewidth]{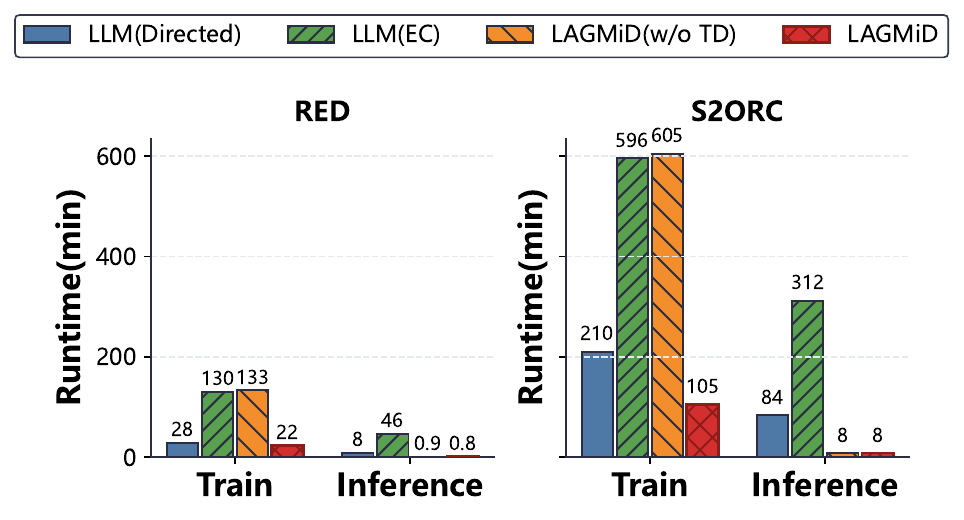}
	\caption{Efficiency comparison of LAGMiD and LLM-based models: Training and inference runtime (in minutes).}
    \Description{ruintime}
	\label{Fig_runtime}
	\vspace*{-5mm}
\end{figure}

To evaluate model efficiency, we compare the runtime of LAGMiD against several LLM-based approaches. As shown in Figure~\ref{Fig_runtime}, LLM (EC), which employs multi-hop reasoning, incurs significantly higher computational cost than the direct reasoning of LLM (Directed). In contrast, both variants of LAGMiD achieve substantial gains in inference speed. Even without uncertainty filtering, LAGMiD (w/o TD) enables rapid inference by distilling LLM reasoning into the GNN across the full training set. The complete LAGMiD model further improves training efficiency through targeted distillation on the uncertainty subset. The complete LAGMiD model further improves training efficiency through targeted distillation on the uncertainty subset, achieving training times close to those of LLM (Directed). During inference, LAGMiD delivers a 10× speedup over LLM (Directed) and a 100× speedup over LLM (EC), demonstrating strong potential for scalable deployment.

%% file: 2RelatedWork.tex
\section{Related Work}
\label{sec:related_work}

\subsection{Miscitation Detection}

Miscitations in scholarly web~\cite{kotiaho1999papers, cobb2024problem, schlichtkrull2023averitec}, also referred to as inaccurate or anomalous citations, occur when cited sources do not substantively support the associated statements. Citation-verification studies consistently find that such errors are common, with prevalence estimates 25\% in the general scientific literature~\cite{peoples2023burden}, underscoring the limits of manual peer review and the need for scalable detection. Existing end-to-end detection approaches fall into two main categories. The first leverages semantic similarity by aligning a citing sentence with candidate argument spans in the referenced paper~\cite{petroni2023improving, wadden2020fact, wadden2022scifact, piccardi2020quantifying}; while this can enforce coarse topical consistency, relatedness does not guarantee semantic adequacy, and fine-grained entailment or reasoning remains challenging. The second mines anomalies in the citation network, for example atypical citation patterns~\cite{ebadulla2025detecting, wren2022detecting,liu2022deep} or signals derived from authors, venues, and other metadata~\cite{liu2024anomalous, kojaku2021detecting}. Such methods can reveal strategic or distorted citation behavior, yet they still cannot directly assess whether a claim is supported by its references. Recent LLMs offer a path to fine-grained, semantics-aware miscitation analysis~\cite{zhang2024detecting}, yet web-scale inference is hindered by hallucinations and prohibitive cost. We address this gap by designing an LLM-based reasoning framework for miscitation detection and distilling its knowledge into a GNN. By combining deep semantic understanding with structure-aware modeling of the citation graph, our approach directly evaluates claim–citation compatibility while retaining the efficiency and scalability required for large-scale deployment.

\subsection{Text-Rich Graph Learning}

Text-rich graphs, also known as text-attributed graphs, arise when nodes and/or edges carry substantial unstructured text~\cite{zhang2021minimally, wang2025text}. Classical approaches analyze such graphs by locally encoding text and propagating signals over the topology~\cite{zou2023pretraining, wu2024hierarchy, wu2025classifying}, or by constructing knowledge graphs whose nodes represent words and documents~\cite{wang2023can}. With the strong text understanding capabilities of LLMs, the integration of LLMs with graphs has become a central direction in text-rich graph learning~\cite{jin2024large, li2024survey, zhou2025data}. Among the LLM-augmented designs most relevant to our setting, three patterns are prominent: LLM-as-Encoder~\cite{wu2024can, fatemi2023talk}, which uses an LLM to produce robust textual embeddings for nodes and edges; LLM-as-Aligner~\cite{wang2025infonce, wang2024llms, liu2024large}, which maps heterogeneous textual and structural signals into a shared space for cross-type reasoning; and retrieval-augmented graph reasoning~\cite{he2024g, peng2024graph, xu2025harnessing}, which constrains or supports LLM judgments using graph neighborhoods. In miscitation detection, textual and structural cues are complementary~\cite{xu2024llm, liu2022deep}. Building on existing paradigms, we propose to distilling the LLM’s reasoning process on the graph into a GNN and couple the distilled knowledge with the GNN’s native structural inference. This design achieves more efficient and scalable text-rich graph learning while preserving fine-grained semantic fidelity and exploiting topology-aware generalization.

%% file: 6Conclusion.tex
\section{Conclusion}
\label{sec:conclusion}
In this paper, we propose LAGMiD, a novel framework that integrates the semantic reasoning power of large language models with graph learning for miscitation detection. By introducing evidence chain reasoning over text-rich citation graphs, distilling such reasoning capacity into a GNN through knowledge transfer, and enabling collaborative optimization with structural features, our method enhances both semantic comprehension of citation contexts and reasoning efficiency. Extensive experiments on three real-world scholarly benchmarks demonstrate that LAGMiD not only achieves state-of-the-art performance in miscitation detection but also significantly accelerates inference compared to LLM-only approaches. 
Looking forward, we believe the proposed framework holds practical potential in supporting academic integrity initiatives on the scholarly web.



%% file: 7Appendix.tex
\section{Construction of Text-Rich Citation Graph from Scholarly Web}
\label{constructionTG}
A systematic preprocessing pipeline is required to transform scholarly web data into a text-rich citation graph $\mathcal{G} = (\mathcal{P}, \mathcal{E}, \mathcal{X}^{(P)}, \mathcal{X}^{(E)})$. This process typically begins with data acquisition from diverse scholarly sources, including academic databases such as Web of Science, Semantic Scholar, and PubMed Central. The nodes $\mathcal{P}$ generally represent individual documents, such as scientific papers or Wikipedia entries. Their corresponding raw text attributes $\mathcal{X}^{(P)}$ can be defined at multiple levels of granularity, including titles, abstracts, and section contents. Edges $\mathcal{E}$ in a text-rich citation graph denote explicit citation relationships between publications. Each directed edge $e_{ij} = (p_i, p_j)$ represents a specific citation where paper $p_i$ references paper $p_j$. Crucially, these edges are also associated with raw text attributes $\mathcal{X}^{(E)}$, which capture the semantic context of the citation relationship. This is achieved by extracting and processing relevant textual information from the citing documents, such as sentences or paragraphs that mention the cited work, and employing advanced natural language processing techniques to identify specific claims or statements about the referenced publication. Such a text-rich citation graph reflects an integrated and interactive characterization of the scholarly web, seamlessly combining structured citation patterns with rich semantic content.

\section{More Implementation Details}

\subsection{Dataset Construction and Processing}
\label{dataset}
\textbf{RED Dataset~\cite{zhang2024detecting}.}  This dataset was originally curated for citation-level fact checking, where each instance is a human-annotated claim–citation pair indicating whether the cited source substantiates the claim or constitutes a reference error. Building on this resource, we augment RED with citation-network structure to enable graph-based learning. Specifically, using Digital Object Identifiers (DOIs) as anchors, we attach bibliographic metadata from Web of Science (WoS) and expand the local citation neighborhood to third-order references, yielding a connected citation graph while preserving the original human-provided labels. For all experiments, we adhere to this augmented RED corpus and partition it into train/dev/test splits in a 7:1:2 ratio to maintain class balance.

\vspace{0.5em} \noindent\textbf{SciFact Dataset~\cite{wadden2020fact}.} This is a scientific-claim verification dataset in which natural‐language claims are paired with evidence from peer-reviewed paper abstracts. Each claim is annotated with a veracity label. In our experiments, we follow the official SciFact data setup: we use the dataset’s prescribed train/dev/test splits without modification, and adopt its label space. This alignment ensures direct comparability with prior work.

\vspace{0.5em}\noindent \textbf{S2ORC Dataset~\cite{lo2020s2orc}.} To address the scarcity of publicly available datasets for miscitation detection, we construct a computer science subset from S2ORC by following the methodology outlined in \cite{liu2024anomalous}. We begin with papers containing complete metadata to build a citation-preserving corpus that maintains authentic references and explicit citing–cited relationships. To simulate miscitation behaviors, we inject synthetically generated anomalous citations into a randomly sampled 50\% of the papers, leaving the remainder unaltered. For each sampled paper, the number of injected anomalies is kept equal to that of its original references, maintaining a 1:1 ratio. The injected anomalies cover three distinct types: (i) citations to collaborators' publications, (ii) within-journal self-promotional citations, and (iii) cross-domain citations that are semantically unrelated to the paper's content.  The corpus is partitioned into training, validation, and test sets in a 7:1:2 ratio.

\begin{table}[t]
\centering
\caption{Performance analysis of evidence chain length $K$.}
\begin{tabular}{lcccc}
\hline
\multirow{2}{*}{Dataset} & \multicolumn{4}{c}{AUC} \\
\cline{2-5}
 & $K=1$ & $K=2$ & $K=3$ & $K=4$ \\
\hline
RED & 0.8362 & \textbf{0.8543} & 0.8470 & 0.8215 \\
S2ORC & 0.7206 & \textbf{0.7680} & 0.7450 & 0.7312 \\
\hline
\end{tabular}
\label{tab:performance-K}
\end{table}

\begin{table}[t]
\centering
\caption{Performance analysis of evidence chain selection parameter $m$ (using Qwen3-8B).}
\vspace*{-3mm}
\label{tab:top_m_analysis}
\begin{tabular}{lcccc}
\hline
\multirow{2}{*}{Dataset} & \multicolumn{4}{c}{AUC} \\
\cline{2-5}
& $m=1$ & $m=5$ & $m=10$ & $m=20$ \\
\hline
RED & 0.8186 & 0.8543 & \textbf{0.8560} & 0.8420 \\
S2ORC & 0.7200 & 0.7480 & \textbf{0.7590} & 0.7460 \\
\hline
\end{tabular}
\end{table}

\begin{table}[t]
\centering
\caption{Performance analysis of uncertainty set ratio $w\%$.}
\vspace*{-3mm}
\label{tab:uncertainty_ratio_analysis}
\begin{adjustbox}{max width=\linewidth}
\begin{tabular}{lccccccc}
\hline
\multirow{2}{*}{Dataset} & \multirow{2}{*}{Metric} & \multicolumn{6}{c}{Uncertainty Set Ratio $w\%$} \\
\cline{3-8}
 & & 1\% & 5\% & 10\% & \textbf{20\%} & 40\% & 50\% \\
\hline
\multirow{4}{*}{RED} & AUC & 0.8600 & 0.9015 & 0.9295 & \textbf{0.9615} & 0.9503 & 0.9455 \\
& F1 & 0.8333 & 0.8511 & 0.8511 & \textbf{0.9167} & 0.9020 & 0.9412 \\
& Precision & 0.8276 & 0.8696 & 0.8889 & \textbf{0.9167} & 0.8519 & 0.8889 \\
\hline
\multirow{4}{*}{S2ORC} & AUC & 0.7429 & 0.7712 & 0.7953 & \textbf{0.8100} & 0.8037 & 0.7989 \\
& F1 & 0.8000 & 0.7879 & 0.8224 & \textbf{0.8256} & 0.8302 & 0.8211 \\
& Precision & 0.7480 & 0.7620 & 0.7787 & \textbf{0.8419} & 0.7700 & 0.7827 \\
\hline
\end{tabular}
\end{adjustbox}
\end{table}

\begin{figure}[t]
	\centering
	\includegraphics[clip=true, width=0.87\linewidth]{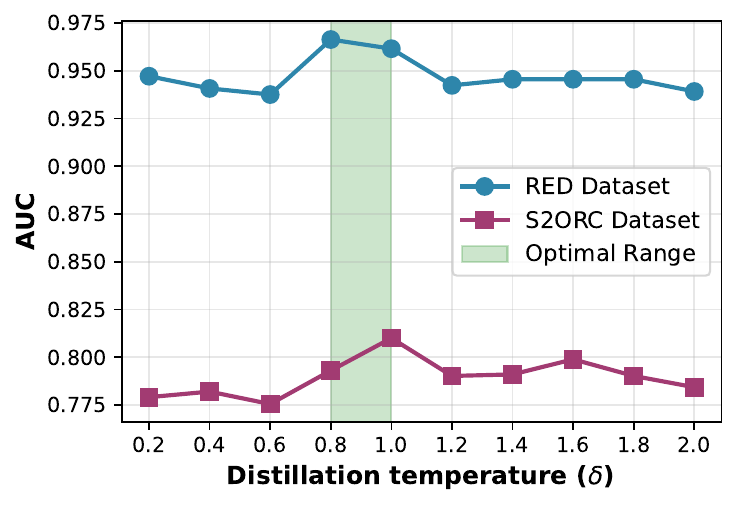}
	\caption{Performance analysis of distillation temperature $\delta$ across datasets.}
    \Description{temperature analysis}
	\label{fig:temperature_analysis}
	\vspace*{-5mm}
\end{figure}

\subsection{Baseline Implementation Details}
\label{app:baseline_details}

We benchmark eight models across four families: GNN based, PLM based, LLM based, and text rich graph learning (TGL). To ensure comparability, we use a unified preprocessing and evaluation pipeline.

\vspace{0.5em}\noindent\textbf{GNN based Methods.}
This category includes GCN and GLAD. Both models use a two layer architecture with a hidden size of 1024, ReLU activations, and dropout of 0.5. We train for 200 epochs with Adam (learning rate \(1\times10^{-3}\)) and use the same graph structure as well as identical node and edge features across methods. 

\vspace{0.5em}\noindent\textbf{PLM based Methods.}
We evaluate RoBERTa and SciBERT. For each instance, we concatenate the claim and the corresponding citation text with a \texttt{[SEP]} delimiter and feed the sequence to the encoder~\cite{petroni2023improving, wadden2020fact}. We use a maximum sequence length of 512, a batch size of 16, and a linearly decayed learning rate of \(2\times10^{-5}\) with AdamW. Inputs are tokenized with the model specific tokenizer.

\vspace{0.5em}\noindent\textbf{LLM based Methods.}
We consider two open source LLMs, namely GLM\textendash 4\textendash 9B and Qwen\textendash 3\textendash 8B, accessed through their official APIs. We standardize decoding with temperature equal to 0.1 and the default \texttt{top\_p} in order to reduce stochasticity. Inputs follow the same claim–citation concatenation scheme with an explicit instruction template for classification. We map generated outputs to discrete labels using fixed regular expressions.

\vspace{0.5em}\noindent\textbf{TGL based Methods.}
From text rich graph learning, we adopt AnomalyLLM and GuARD. Both models operate on text attributed graphs for anomaly detection at the edge level. We treat anomalous edges as miscitations and train edge level detectors accordingly. Following the original recommendations, we construct text features from the claim and citation content using the same PLM encoder across baselines, retain identical graph connectivity, and use the default losses and regularizers for graph based anomaly detection.

\begin{figure*}[htbp]
	\centering
	\includegraphics[clip=true, trim= 30 0 20 0, width=0.87\linewidth]{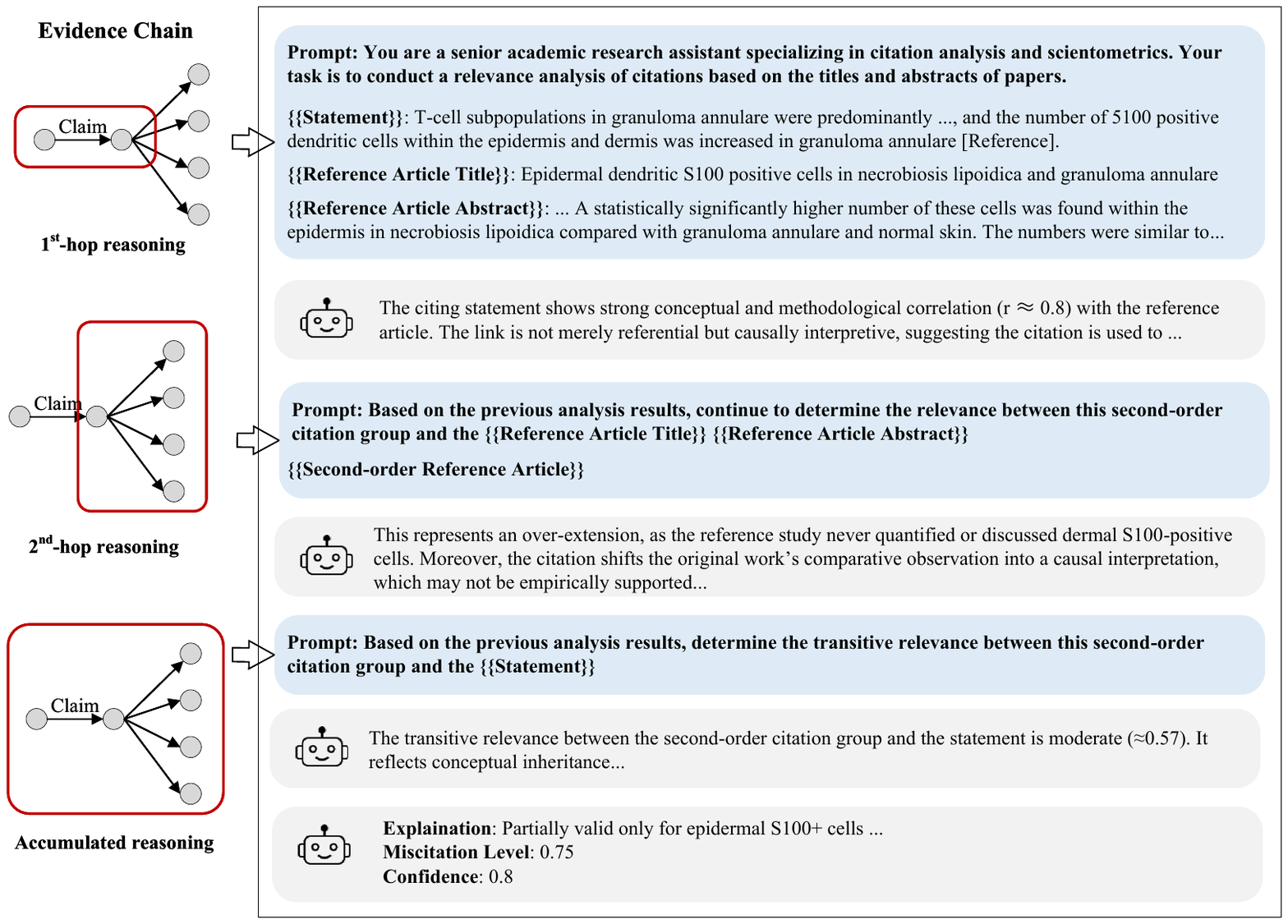}
    \vspace*{-5mm}
	\caption{Case study of a 2-hop evidence-chain reasoning process with Chain-of-Thought.}
    \Description{A case of Evidence-Chain Reasoning}
	\label{fig:case}
	\vspace*{-2mm}
\end{figure*}

\section{Parameter Analysis}

Our approach incorporates several hyperparameters, and we present a detailed analysis of four key ones.

\vspace{0.5em} \noindent \textbf{Evidence Chain Length Parameter $K$.} The chain-of-thought length $K$ determines the number of reasoning hops. As shown in Table~\ref{tab:performance-K}, when $K=2$, the model achieves the best performance on both the RED and S2ORC datasets. When $K=1$, the model's performance is limited due to its inability to leverage multi-hop dependencies; when $K>2$, excessively long reasoning chains introduce noise and increase computational overhead, leading to performance degradation. Therefore, this paper sets $K=2$ to achieve the optimal balance between capturing complex citation relationships and maintaining reasoning efficiency.

\vspace{0.5em} \noindent \textbf{Evidence Chain Selection Parameter $m$.} The parameter $m$ determines the number of top relevant nodes retained at each hop during evidence chain extraction. As summarized in Table~\ref{tab:top_m_analysis}, both datasets exhibit similar trends: performance initially improves with larger $m$, peaks at moderate values ($m = 5$--$10$), and slightly declines when $m$ becomes too large ($m = 20$). This pattern suggests that selecting too few nodes ($m = 1$) may fail to capture valuable citation sources across multiple hops, while an excessively broad selection ($m = 20$) introduces noise that undermines model performance.

\vspace{0.5em} \noindent \textbf {Uncertainty Set Ratio $w\%$.}
The uncertainty set ratio $w\%$ specifies the proportion of uncertain edges selected for refinement by a large language model (LLM) in our collaborative learning framework. Table~\ref{tab:uncertainty_ratio_analysis} provides comprehensive results across different ratios. We observe that both datasets achieve optimal performance at $w\% = 20\%$, with RED attaining a peak AUC of 0.9615 and S2ORC reaching 0.8100. This optimal value reflects a trade-off between computational efficiency and reasoning quality: smaller ratios ($w\% < 20\%$) risk overlooking informative uncertain instances, while larger ratios ($w\% > 20\%$) incur additional computational cost without yielding meaningful performance improvements.

\vspace{0.5em} \noindent \textbf {Distillation Temperature Parameter $\delta$.}
The temperature parameter $\delta$ in the InfoNCE loss controls the concentration level of the knowledge distribution. As illustrated in Figure~\ref{fig:temperature_analysis}, model performance remains robust across a range of $\delta$ values. The optimal range of $\delta = 0.8$--$1.0$ indicates that moderate temperature settings strike an effective balance between exploration and exploitation in the knowledge distillation process. Excessively low temperatures ($\delta < 0.8$) may produce over-confident distributions that impede knowledge transfer, while overly high values ($\delta > 1.0$) yield excessively smooth distributions.

\section{Case Study}
\label{case_study}
Figure~\ref{fig:case} provides a specific example of our LLM-based evidence-chain reasoning. This case showcases a 2-hop chain-of-thought reasoning procedure. The left side of the figure depicts a schematic of the extracted evidence chain, with red boxes highlighting the scope of reasoning focus at each step. The right side details the corresponding prompt template and the intermediate reasoning outputs generated by the LLM. The process begins with the LLM analyzing the semantic alignment between the initial claim and its direct citation. Subsequently, the model evaluates the validity of the citation in relation to its own 2-hop citation sources. Finally, by synthesizing the multi-step reasoning trajectory, the LLM assesses the overall logical coherence of the claim within the citation chain and determines the degree of miscitation.